\documentclass[published]{JHEP3} 
\JHEP{12(2003)013}
\received{October 7, 2003}
\accepted{December 10, 2003}

\usepackage{multicol}

\newcommand\SU{\mathop{\rm SU}\nolimits} 

\title{Noncommutative field theory and violation of translation invariance}

\author{Orfeu Bertolami and Lu\'\i s Guisado\thanks{Deceased.}\\ 
Departamento de F\'\i sica, Instituto Superior T\'ecnico\\
Av. Rovisco Pais 1, 1049-001 Lisboa, Portugal\\
E-mail: \email{orfeu@cosmos.ist.utl.pt}}

\abstract{Noncommutative field theories with commutator of the
  coordinates of the form $[x^{\mu },x^{\nu }]=i\, \Lambda_{\quad
    \omega}^{\mu \nu }x^{\omega }$ with nilpotent structure constants
  are studied and shown that a free quantum field theory is not
  affected. Invariance under translations is broken and the
  conservation of energy-momentum is violated, obeying a new law which
  is expressed by a Poincar\'e-invariant equation. The resulting new
  kinematics is studied and applied to simple examples and to
  astrophysical puzzles, such as the observed violation of the GZK
  cutoff.

  The $\lambda $$\Phi ^{4}$ quantum field theory is also considered in
  this context. In particular, self interaction terms violate the
  usual conservation of energy-momentum and, hence, the radiative
  correction to the propagator is altered. The correction to first
  order in $\lambda $ is calculated. The usual UV divergent terms are
  still present, but a new type of term also emerges, which is IR
  divergent, violates momentum conservation and implies a correction
  to the dispersion relation.}

\dedicated{Dedicated to the memory of Lu\'\i s Guisado.}

\begin{document}

\section{Introduction}

Noncommutativity of coordinates has been intensively studied in the
literature as it arises in the context of string
theory~\cite{Seiberg_Witten}, but also because it has interesting
properties and implications in field
theory~\cite{Szabo,Doug_Nekra}. In many treatments of noncommutative
field theory the noncommutative parameters are not regarded as Lorentz
tensors, but instead a set of numbers that do not transform
covariantly which implies naturally in the breaking of Lorentz
invariance down to the stability subgroup of the noncommutative
parameter~\cite{Carroll}. Furthermore, noncommutative structures of
the Lie-type imply in the violation of the energy-momentum
conservation as the coordinate commutation relations break
translational invariance.  Alternatively, one could consider instead
the noncommutative parameter as a Lorentz tensor covariant under
Lorentz boosts. This approach has been studied earlier and in this
framework we have shown that a noncommmutative scalar field coupled to
gravity admits a covariant formulation (where
associativity\footnote{The relation between a nonassociative star 
product on D-branes and noncommutative theories on curved spaces 
has been discussed in~\cite{Cornalba}.}
 is mantained only at perturbative level)
which is compatible with a homogeneous and isotropic
space-time~\cite{Bertolami_Guisado}.  Further attempts along these
lines include work on noncommutative scalar field theory in
three-dimensions~\cite{Sasakura} and on QED~\cite{Conroy}.  We mention
that in the latter, gauge invariance is implemented via the
introduction of a noncommutative gauge field, a ``star'' gauge
invariance and ``star'' commutators.

\pagebreak[3]

In this work we shall analyse classical and quantum field theory
features of models where the noncommutativity of the coordinates has
the following form
\begin{equation}
\left[x^{\mu },x^{\nu }\right]=i\, \Lambda _{\quad \omega }^{\mu \nu }
x^{\omega }\,,
\label{intro}
\end{equation}
with the condition of nilpotency as specified below.  This leads to a
violation of the symmetry under translations and, consequently,
requires a reformulation of energy-momentum conservation.  This
reformulation is proven to be Poincar\'e-invariant and reduces to the
usual momentum conservation in the commutative limit. The formalism we
develop follows the study of ref.~\cite{Kathotia}, where the
Baker-Hausdorff formula is related to the Kontsevich
(see~\cite{Kontsevich} and ealier references therein) noncommutative
product, for Lie algebras of the form~(\ref{intro}).

Before closing our introduction, let us point out that our nilpotency
condition (cf.\ eq.~(\ref{nilpotency}) below) excludes noncommutative
structures of the Lie-type such as the semisimple Lie algebras
($\SU(2)$ for the fuzzy sphere) and the $\kappa$-deformed Minkowsky
space that are often discussed in the literature (see
ref.~\cite{Kosinski} for an extensive review).  Furthermore, we
mention that recent work by Robbins and Sethi~\cite{Robbins} is
closely related with ours, even though it considers examples that are
more directly inspired by string theory.

\section{Mathematical formulation}

\subsection{Noncommutative algebra}

A noncommutative associative product may be defined through the
Lie-algebra commutator eq.~(\ref{intro}), where $\Lambda ^{\mu \nu
  \omega }$ is a real tensor with units of $\textrm{mass}^{-1}$ and
$\Lambda ^{\mu \nu \omega }=-\Lambda ^{\nu \mu \omega }$.  On its
hand, associativity implies the Jacobi identity
\begin{equation}
\Lambda _{\quad \omega }^{\mu \nu }
\Lambda _{\quad \beta }^{\omega \alpha }+
\Lambda _{\quad \omega }^{\nu \alpha }
\Lambda _{\quad \beta }^{\omega \mu }+
\Lambda _{\quad \omega }^{\alpha \mu }
\Lambda _{\quad \beta }^{\omega \nu }=0\,.
\label{Jacobi}
\end{equation}

A noncommutative Fourier mode is defined by
\begin{equation}
e_{*}^{ik\cdot x}=\sum _{n=0}^{\infty }{i^{n} \over n!}
\overbrace{\left(k\cdot x\right)*\cdots *
\left(k\cdot x\right)}^{n\textrm{ factors}}=
\sum _{n=0}^{\infty }{i^{n} \over n!}\left(k\cdot x\right)_{*}^{n}\,,
\label{Fourier}
\end{equation}
and we study the functional space spanned by these Fourier modes, with
elements of the form
\begin{equation}
f\left(x\right)=\int {d^{n}k \over \left(2\pi \right)^{n}}\tilde{f}
\left(k\right)e_{*}^{ik\cdot x}
\label{NC_Fourier}
\end{equation}
which, in the commutative limit, reduces to the usual Hilbert space.
Notice that in eq.~(\ref{Fourier}) the star product acts only on the
configuration variables and not on the momentum ones.

The product of two generic functions is then given by
\begin{equation}
f*g=\int {d^{n}k \over \left(2\pi \right)^{n}} {d^{n}q \over  \left(2\pi
  \right)^{n}}\tilde{f}\left(k\right)\tilde{g}\left(q\right)e_{*}^{ik\cdot
  x}*e_{*}^{iq\cdot x}\,,
\end{equation} 
where we have expressed the functions in terms of their noncomutative
Fourier expansion. This product is completely determined if the
product of two Fourier modes $e_{*}^{ik\cdot x}*e_{*}^{iq\cdot x}$ can
be evaluated. This can be achieved by making use of the
Baker-Hausdorff formula
\begin{equation}
  e_{*}^{ik\cdot x}*e_{*}^{iq\cdot x}=\exp _{*}\left\{
  i\left(k+q\right)\cdot x+{1 \over 2}\left[ik\cdot x,iq\cdot
    x\right]+\cdots \right\} ,
\end{equation} 
where the dots stand for higher order commutators. Since the
commutators obey
\begin{equation}
\left[x^{\mu _{1}},\left[x^{\mu _{2}},\ldots ,\left[x^{\mu
	_{n}},x^{\nu }\right]\right]\cdots \right]\propto
i^{n}x^{\omega }\,,
\end{equation} 
the product of two Fourier modes is a Fourier mode
\begin{equation} 
e_{*}^{ik\cdot x}*e_{*}^{iq\cdot
  x}=e_{*}^{i\left[k+q+V\left(k,q\right)\right]\cdot x}
\end{equation}
with $V$ determined by the Baker-Hausdorff expansion:
\begin{equation}
  V_{\omega }\left(k,q\right)=k_{\mu }q_{\nu }\Lambda _{\quad \lambda
  }^{\mu \nu }\left[-{1 \over 2}\delta _{\omega }^{\lambda }
    +{k_{\alpha }-q_{\alpha } \over 12}\Lambda _{\quad \omega
    }^{\alpha \lambda }\right]+{O(\Lambda ^{3})}\,  .
\end{equation}

\subsection{Quadratic actions}

In order to build actions, a star-integration must be defined. In the
functional space whose elements are of the form~(\ref{NC_Fourier}),
any function can be integrated if the integral of a Fourier mode is
known. Hence, we introduce the following star-integration
\begin{equation}
\int _{*}d^{n}x\, e_{*}^{ir\cdot x}=\left(2\pi \right)^{n}\delta 
\left(r\right),
\end{equation}
which yields the usual integration in the commutative limit.

Consider now the star-integral
\begin{equation}
I=\int _{*}d^{n}x\, f*g\: .
\end{equation}
which, in Fourier space, is written as
\begin{equation}
I=\int {d^{n}k \over \left(2\pi \right)^{n}}{d^{n}q \over \left(2\pi
  \right)^{n}}\tilde{f}\left(k\right)\tilde{g}\left(q-k\right)\int
_{*}d^{n}x\, e_{*}^{i\left[q+V\left(k,q-k\right)\right]\cdot x}\,,
\end{equation}
and implies
\begin{equation}
I=\int {d^{n}k \over \left(2\pi
  \right)^{n}}d^{n}q\tilde{f}\left(k\right)\tilde{g}\left(q-k\right)\delta
\left(q+V\left(k,q-k\right)\right).
\end{equation}

If the structure constants are nilpotent, that is, for $n>n_{*}$
\begin{equation}
\Lambda _{\quad \, \omega _{1}}^{\mu _{1}\nu }
\Lambda _{\quad \quad \omega _{2}}^{\mu _{2}\omega _{1}}\cdots 
\Lambda _{\qquad \quad \omega _{n}}^{\mu _{n}\omega _{n-1}}=0\, ,
\label{nilpotency}
\end{equation}
then
\begin{equation}
\delta \left(q+V\left(k,q-k\right)\right)=
{\delta \left(q\right) \over \left|\det \left(\delta _{\nu }^{\mu }-
{\partial V_{\nu } \over \partial q_{\mu }}\right)\right|}=
\delta \left(q\right)
\end{equation}
since $\det \left(1+M\right)=1$ if $M^{n}=0$, which holds if $\Lambda
$ is nilpotent. Thus
\begin{equation}
I=\int {d^{n}k \over \left(2\pi \right)^{n}}\tilde{f}\left(k\right)
\tilde{g}\left(-k\right)=\int d^{n}x\, f_{C}\left(x\right)g_{C}
\left(x\right)
\label{quadratic}
\end{equation}
where $f_{C}$, $g_{C}$ are inverse Fourier transforms \emph{using
  commutative Fourier modes}
\begin{equation}
f_{C}\left(x\right)=\int {d^{n}k \over \left(2\pi \right)^{n}}
\tilde{f}\left(k\right)e^{ik\cdot x}\,.
\end{equation}
Equation~(\ref{quadratic}) states that, in momentum space, quadratic
terms in the lagrangian are the same as their commutative
counterparts.  In particular, free propagators will remain unchanged.

\section{Violation of momentum conservation}

We have concluded that the quadratic part of a lagrangian is not
changed and, hence, the free theory is the same as the commutative
one. In particular, the free Green function is equal to the
commutative case and the dispersion relation $\epsilon
^{2}=p^{2}+m^{2}$ is unchanged, since it is given by the poles of the
free propagator. Yet, we shall see that interactions are altered by
non-commutativity.

Consider a non-commutative field theory, with generic fields $A_{i}$
and an interaction term 
\begin{equation} 
S_{I}=\int _{*}d^{n}x\,
  M_{i_{1}\ldots i_{m}}A_{i_{1}}*\cdots *A_{i_{m}}\, ,
\end{equation}
where $M_{i_{1}\cdots i_{m}}$ are constants.

Writing the fields in momentum space we get
\begin{equation}
S_{I}=\int \left[\prod _{i=1}^{m}{d^{n}k_{i} \over 
\left(2\pi \right)^{n}}\right]\tilde{M}_{i_{1}\cdots i_{m}}
\left(\underline{k}_{m}\right)\tilde{A}_{i_{1}}\left(k_{1}\right)
\cdots \tilde{A}_{i_{m}}\left(k_{m}\right)
\label{mute}
\end{equation}
where we use the notation $\underline{k}_{m}=\left(k_{1},\ldots
,k_{m}\right)$.  The interaction in momentum space is given~by
\begin{equation}
\tilde{M}_{i_{1}\cdots i_{m}}\left(\underline{k}_{m}\right)=
M_{i_{1}\cdots i_{m}}\int _{*}d^{n}x\, e_{*}^{ik_{1}\cdot x}*\cdots *
e_{*}^{ik_{m}\cdot x}\,.
\label{new_vertex}
\end{equation}

In eq.~(\ref{mute}) the variables $k_{i}$ are mute, so we can sum over
all $\pi $ permutations of the indices $i_{m}$:
\begin{equation}
S_{I}=\int \left[\prod _{i=1}^{m}{d^{n}k_{i} \over \left(2\pi
    \right)^{n}}\right]\tilde{M}_{i_{1}\cdots
  i_{m}}^{\rm symm}\left(\underline{k}_{m}\right)\tilde{A}_{i_{1}}
\left(k_{1}\right)\ldots
\tilde{A}_{i_{m}}\left(k_{m}\right),
\end{equation}
where
\begin{equation}
\tilde{M}_{i_{1}\cdots i_{m}}^{\rm symm}\left(\underline{k}_{m}\right)={1
  \over m!}\sum _{\pi \, {\rm perm.}}\left(-\right)^{N\left(\pi
  \right)}\tilde{M}_{i_{\pi \left(1\right)}\cdots i_{\pi
    \left(m\right)}}\left(\underline{k}_{\pi \left(m\right)}\right).
\end{equation}

To evaluate eq.~(\ref{new_vertex}), we use the expression
\begin{equation}
e_{*}^{ik_{1}\cdot x}*\cdots *e_{*}^{ik_{m}\cdot x}=\exp _{*}\left\{ 
i\sum _{j=1}^{m}k_{j}\cdot x+iV^{m}\left(\underline{k}_{m}\right)
\cdot x\right\} 
\end{equation}
where 
\begin{equation} 
V^{m}\left(\underline{k}_{m}\right)=V^{m-1}\left(\underline{k}_{m-1}\right)
+V\left(\sum _{i=1}^{m-1}k_{i}+V^{m-1}\left(\underline{k}_{m-1}\right),k_{m}
\right)
\end{equation}
with $V^{2}\left(\underline{k}_{2}\right)=V\left(k_{1},k_{2}\right)$.
This yields both the noncommutative energy-momentum law
and the noncommutative vertex
\begin{equation}
\tilde{M}_{i_{1}\cdots i_{m}}\left(\underline{k}_{m}\right)=
\left(2\pi \right)^{n}\delta \left(\sum _{i=1}^{m}k_{i}+V^{m}
\left(\underline{k}_{m}\right)\right)M_{i_{1}\cdots i_{m}}\,.
\label{nc_vertex}
\end{equation}

Hence, the new energy-momentum law for the vertex reads
\begin{equation}
\sum _{i=1}^{m}k_{i}+V^{m}\left(\underline{k}_{m}\right)=0\,.
\end{equation}

The full theory involves $\tilde{M}_{i_{1}\cdots i_{m}}^{symm}$ , which
will have contributions whenever 
\begin{equation}
\sum _{i=1}^{m}k_{i}+V^{m}\left(\underline{k}_{\pi \left(m\right)}\right)=0
\label{cons_momentum}
\end{equation}
for all $m!$ permutations of indices, $\pi $.

Thus, we see that the energy-momentum conservation is violated as the
theory is not invariant under translations. In fact, in a translation
$x^{\mu }\rightarrow x^{\mu }+b^{\mu }$, the commutator of the
coordinates is changed by
\begin{equation}
\left[x^{\mu },x^{\nu }\right]\rightarrow i\, 
\Lambda _{\quad \omega }^{\mu \nu }x^{\omega }+i\, \theta ^{\mu \nu }\, ,
\end{equation}
that is, a constant term $\theta ^{\mu \nu }=\Lambda _{\quad \omega
}^{\mu \nu }b^{\omega }$ is added to the commutator of the
coordinates. So, the interaction vertex becomes
\begin{equation} 
\tilde{M}_{i_{1}\cdots i_{m}}\left(\underline{k}_{m}\right)\rightarrow
\left(2\pi \right)^{n}\delta \left(\sum
_{i=1}^{m}k_{i}+V^{m}\left(\underline{k}_{\pi
  \left(m\right)}\right)\right)M_{i_{1}\cdots i_{m}}\exp \left\{
i\theta ^{m}\left(\underline{k}_{m}\right)\right\}
\end{equation}
where
\begin{equation}
\theta ^{m}\left(\underline{k}_{m}\right)=\theta
^{m-1}\left(\underline{k}_{m-1}\right)+\theta \left(\sum
_{i=1}^{m-1}k_{i}+V^{m-1}\left(\underline{k}_{m-1}\right),k_{m}\right)
\end{equation}
and $\theta ^{2}\left(\underline{k}_{2}\right)=k_{1\mu }\theta ^{\mu
  \nu }k_{2\nu }$.  Hence, the interaction vertex is altered by an
overall oscillating momentum-dependent factor and, thus, invariance
under translations is broken. This example shows that translations
give always rise to a constant term in the noncomutative
tensor. However, the new energy- momentum law is unchanged, so it is a
Poincar\'e-invariant expression, even though the theory is not.

\section{Kinematical applications}

\subsection{Preliminaries}

The first non-trivial behaviour arising from the new interaction vertex
occurs with three particles. The energy-momentum equation
is found to be
\begin{equation}
k_{1}+k_{2}+k_{3}+V\left(k_{1},k_{2}\right)+V\left(k_{1}+k_{2}
+V\left(k_{1},k_{2}\right),k_{3}\right)=0
\label{cons_decay}
\end{equation}
and similar expressions for all permutations of the indices.

\pagebreak[3] 

On physical grounds, eq.~(\ref{cons_decay}) represents three particles
interacting. Formally, these equations can be treated in the context
of the usual momentum conservation if one thinks in terms of four
interacting particles, with the fourth particle's energy-momentum
vector being given by a nonlinear function of the others. This
reasoning may be extended to the $m$-particle case,
eq.~(\ref{cons_momentum}).

For the time being, let us consider a simple model with
\begin{equation}
\Lambda _{\quad \omega _{1}}^{\mu _{1}\nu }\Lambda _{\quad \omega
  _{2}}^{\mu _{2}\omega _{1}}=0
\label{nil2}
\end{equation}
which complies with the Jacobi identity eq.~(\ref{Jacobi}).

The energy-momentum equation becomes
\begin{equation}
k_{1}+k_{2}+k_{3}+V\left(k_{1},k_{2}\right)=0\, ,
\label{cons_decay_simple}
\end{equation}
where
\begin{equation}
V_{\omega }\left(k_{1},k_{2}\right)={1 \over 2}k_{1\mu }k_{2\nu
}\Lambda _{\quad \omega }^{\mu \nu }\, .
\end{equation}

There are nontrivial covariant solutions to eq.~(\ref{nil2}). For
instance, consider a constant antisymmetric tensor $\Lambda ^{\mu \nu
}=-\Lambda ^{\nu \mu }$ with nontrivial kernel, that is, $\det \Lambda
=0$, and a non-vanishing vector $r^{\omega }$ belonging to this
kernel. Hence a solution is given by
\begin{equation}
\Lambda ^{\mu \nu \omega }=\Lambda ^{\mu \nu }r^{\omega }\, .
\label{solution}
\end{equation}
In four dimensions we can parametrize $\Lambda ^{\mu \nu }$ with two
spatial vectors $\vec{E}$ and $\vec{B}$
\begin{equation}
\Lambda ^{\mu \nu }=\pmatrix{
 0 & E_{x} & E_{y} & E_{z}\cr
 -E_{x} & 0 & -B_{z} & B_{y}\cr
 -E_{y} & B_{z} & 0 & -B_{x}\cr
 -E_{z} & -B_{y} & B_{x} & 0},
\qquad 
r_{\nu }=\pmatrix{
 r_{0}\cr
 r_{x}\cr
 r_{y}\cr
 r_{z}}.
\end{equation}
Condition eq.~(\ref{solution}) implies that
\begin{equation}
r^{2}=\left|\vec{r}\right|^{2}\left[\left({B \over E}\sin \delta
  \right)^{2}-1\right],
\label{tachyon?}
\end{equation}
with $\delta $ being the angle between $\vec{B}$ and $\vec{r}$.  The
massless, massive and tachyon regimes of $V$ are readily identifiable.
Since we assume that $\Lambda ^{\mu \nu }$ is a Lorentz tensor, there
are always inertial frames where $\vec{E}$ is non-vanishing, and the
above expression holds only for such frames. If $B<E$ (a
Lorentz-invariant inequality) then $r^{\omega }$ behaves like a
tachyon; otherwise, the behaviour of $r^{\omega }$ will depend on
$\delta $.

From the momentum conservation, eq.~(\ref{cons_decay_simple}), we get
the following result
\begin{equation}
\Lambda ^{\mu \nu }\left(k_{1}+k_{2}+k_{3}\right)_{\nu }=0\, ,
\label{kernel_cons}
\end{equation}
which states that the vector sum of the momenta belongs to the
(nontrivial) kernel of the noncommutative tensor. We also have the
following expressions
\begin{equation}
\Lambda ^{\mu \nu }k_{\nu }=
\pmatrix{
\vec{E}\cdot \vec{k}\cr
 -k_{0}\vec{E}+\vec{B}\times \vec{k}}
\label{lambda_k}
\end{equation}
and
\begin{equation}
q_{\mu }\Lambda ^{\mu \nu }k_{\nu }=\vec{E}\cdot \left(q_{0}\vec{k}-k_{0}
\vec{q}\right)
+\vec{B}\cdot \left(\vec{k}\times \vec{q}\right).
\end{equation}

Eqs.~(\ref{kernel_cons}) and~(\ref{lambda_k}) imply that the
three-momentum is conserved along the direction of $\vec{E}$. Energy
is conserved if the total three-momentum $\sum \vec{k_{i}}$ is along
the direction of $\vec{B}$. Also,
\begin{equation} 
k_{1}\Lambda
  k_{2}=-k_{1}\Lambda k_{3}=k_{2}\Lambda k_{3}
\end{equation} 
and one is required only to study eq.~(\ref{cons_decay_simple}) with
$V\left(k_{1},k_{2}\right)$ and $-V\left(k_{1},k_{2}\right)$. Note
that the second case is obtained by performing
$\vec{E},\vec{B}\rightarrow -\vec{E},-\vec{B}$.  Thus, once
computations have been performed in the first case, the results in the
second one are obtained by performing this substitution.

While performing calculations, the following dimensionless combinations
of the masses and the noncommutative parameters arise:
\begin{equation}
x_{i}={1 \over 2}E\left|\vec{r}\right|m_{i}\,, 
\qquad 
y_{i}={1 \over 2}Bs_{\delta }\left|\vec{r}\right|m_{i}\,.
\end{equation}
where the notation $s_{\omega }=\sin \omega $ is used.

\subsection{Massive particle decay}

Consider now the decay of a massive particle $\Phi _{3}$ into two
particles $\Phi _{1}$ and $\Phi _{2}$, that is
\begin{equation}
\Phi _{3}\rightarrow \Phi _{1}+\Phi _{2}\, .
\end{equation}

Let $m_{i}$ be the mass of particle $\Phi _{i}$ and $m_{1}\geq m_{2}$.
In the rest-frame of $\Phi _{3}$ the angle $\alpha $ between particles
$\Phi _{1}$ and $\Phi _{2}$ is given by
\begin{equation}
\cos\alpha =-{1+x_{3}c_{\theta }c_{\varphi } \over
  \sqrt{(1+x_{3}c_{\theta }c_{\varphi
    })^{2}+(x_{3}c_{\theta }s_{\varphi})^{2}}}
\end{equation}
where $\theta $ is the angle between $\vec{p}_{1}$ and $\vec{E}$, and
$\varphi $ the angle between $\vec{p}_{1}$ and $\vec{r}$. Also, the
notation $c_{\omega }=\cos \omega $ is used. The absolute value of the
right-hand side of this equation is always smaller than one, meaning
that this decay is always possible. In the high energy regime ($x_{3}
\gg 1$) the variable $x_{3}$ decouples and one obtains $\alpha \approx
\pm \varphi $ or $\alpha \approx \pi \pm \varphi $. In the low-energy
regime ($x_{3} \ll 1$) one finds the first-order correction $\alpha
\approx \pi \pm x_{3}s_{\varphi }c_{\theta }$.

The new equation for the energy is
\begin{equation}
m_{3}=\epsilon _{1}\left(1-y_{3}c_{\theta }v_{1}\right)+\epsilon _{2}\,,
\end{equation}
where we have used $\epsilon _{i},v_{i}$ as the energy and velocity of
particle $i$.

If $\left|y_{i}\right|<1$ then spontaneous decay will occur if $m_{3}$
satisfies the condition
\begin{equation}
m_{3}>{{m_{1}+m_{2} \over 1+y_{1}c_{\theta }v_{1}}}\, .
\end{equation}
If the denominator is zero or negative, then the decay is impossible.
We can see that if the velocity of particle 1 is high and mainly along
the direction of $\vec{E}$ then the decay can occur for a value of
$m_{3}$ smaller than the sum of the masses $m_{1}+m_{2}$
\begin{equation}
m_{3}>{{m_{1}+m_{2} \over 1+|y_{1}|}}\, .
\end{equation}

\subsection{Massless particle decay}

The decay of a massless particle into two massive particles is
kinematically forbidden. However, in the present model, all massless
particles become unstable and can decay into two massive particles. If
the particle $\Phi _{3}$ is massless we have the following condition
for the energy of the photon in the rest frame of particle $\Phi _{1}$
\begin{equation}
\omega >{2 \over |Bs_{\delta }-Ec_{\varphi
  }||\vec{r}||c_{\theta }|} \equiv {\omega
  _{0}(\varphi ) \over |c_{\theta }|}\, ,
\end{equation}
where we have defined $\theta $ as the angle between $\vec{E}$ and
$\vec{\omega }$ and also $\varphi $ as the angle between $\vec{r}$ and
$\vec{\omega }$. If the denominator is zero, the decay is impossible.
Note that this limit is independent of the mass of the decaying
particles.  This result is only valid in the low-energy limit where
$\left|x_{1}\right|<1$, that is, when the decay produces particles
with low mass.

The above limit for angle $\theta $ implies that
\begin{equation}
c_{\theta }^{2}>\left({\omega _{0} \over \omega }\right)^{2},
\end{equation}
which states that, as the energy of the photon grows larger, the decay
is possible for a wider range of $\theta $ angles. If $\theta =0$, the
decay is impossible.

\subsection{The GZK cutoff}

The Greisen-Zatsepin-Kuz'min (GZK) cutoff mechanism asserts that
ultra-high-energy (UHE) protons with energies $\epsilon _{p}>4\times
10^{19}\, eV$ from sources beyond $50-100\, Mpc$ should not be
observed, due to their interaction with Cosmic Microwave Background
(CMB) photons.  It has been proposed (for brief review see
ref.~\cite{Bertolami}) that Lorentz-violating terms in the kinematics
of hadronic reactions may be the answer to this puzzle. The GZK cutoff
has the following dominant resonance
\begin{equation}
p+\gamma _{CMB}\rightarrow \Delta _{1232}\, .
\label{GZK}
\end{equation}

It is easily shown that the model eq.~(\ref{solution}) does not
account for a violation of the GZK cutoff. In fact, in the case of
head-on collision, the new equation for the energy yields
\begin{equation}
\epsilon _{p}[1-y_{\omega }c_{\theta}
(1+v_{p})]+\omega =\epsilon _{\Delta }\, ,
\end{equation}
where $\epsilon _{i},v_{i}$ denote the energy and velocity of particle
$i$ and $\theta $ is the angle between $\vec{E}$ and $\vec{\omega }$.
In order to occur any appreciable deviation that renders this reaction
impossible (for instance $\epsilon _{\Delta }<m_{\Delta }$), one
should have $y_{\omega }\approx 1$, which, given the low energy of the
$CMB$ photon, would yield a very small mass for the noncommutative
parameters.

Nevertheless, the violation of the GZK limit may be explained in the
context of the model
\begin{equation}
\Lambda _{\quad \omega _{1}}^{\mu _{1}\nu }
\Lambda _{\qquad \omega _{2}}^{\mu _{2}\omega _{1}}
\Lambda _{\qquad \omega _{3}}^{\mu _{3}\omega _{2}}=0\, ,
\end{equation}
with $\Lambda _{\quad \omega _{1}}^{\mu _{1}\nu }\Lambda _{\quad
  \omega _{2}}^{\mu _{2}\omega _{1}}\neq 0$.  This cannot be
implemented by model eq.~(\ref{solution}), which complicates the
analysis. The equation for the momentum is given by
\begin{equation}
\left(k_{1}+k_{2}+k_{3}\right)_{\omega }=k_{1\mu }k_{2\nu }\Lambda
_{\quad \lambda }^{\mu \nu }\left[-{1 \over 2}\delta _{\omega
  }^{\lambda }+{\left(k_{1}-k_{2}\right)_{\alpha } \over 12}\Lambda
  _{\quad \omega }^{\alpha \lambda }\right]
\label{GZK_v2}
\end{equation}
where we have used eq.~(\ref{cons_decay}) recursively and the fact
that cubic terms in $\Lambda $ vanish.

\pagebreak[3] 

This condition can be modeled by a simpler one, more suitable for
phenomenological considerations which, however, breaks Lorentz
invariance.  As we have seen, the quadratic term in the momentum does
not account for the violation of the GZK cutoff, so it will be
dropped. Taking into account that the proton has the highest energy
and the $\Delta $ the second highest energy, we can write the new
momentum equation for the reaction~(\ref{GZK}) as
\begin{equation}
\left(k_{p}+k_{\gamma }\right)^{\mu }=k_{\Delta }^{\mu }-s^{\mu
}{\epsilon _{p}^{2} \over M^{2}}\epsilon _{\Delta }
\label{GZK_funciona}
\end{equation}
where adimensional vector $s^{\mu }$ is of the order of unity and $M$
is the typical noncommutative mass scale. In this case, the process is
impossible if $s^{0}>0$ and $\epsilon _{p}>M$, which sets the scale of
noncommutativity.

Note that one must consider all permutations of the indices in
eq.~(\ref{GZK_v2}). Due to the low energy of CMB photons, the
permutations that lead to a term of the type $\epsilon _{\gamma }^{2}$
will not violate the GZK cutoff. Since there are six permutations of
the indices and only two lead to this type of term, we can estimate
that $2/3$ of the events leading to the ressonance~(\ref{GZK}) will
violate the GZK cutoff.

It is generally believed~\cite{Bertolami,Amelino} that a cubic term in
the equations of dispersion will explain the violation of this
cutoff. In fact, eq.~(\ref{GZK_funciona}) can be obtained by assuming
the usual momentum conservation and postulating a new equation of
dispersion by the substitution
\begin{equation}
k^{\mu }\rightarrow k^{\mu }+s^{\mu }{\epsilon ^{2} \over M^{2}}\lambda 
\end{equation}
where $\lambda $ represents the typical energy of the product of the
reaction. This will lead to the following dispersion relation
\begin{equation}
m^{2}=\epsilon ^{2}-p^{2}+2s^{\mu }v_{\mu }{\lambda \over M^{2}}\epsilon ^{3}
\label{cubic_eq_disp}
\end{equation}
where $v^{\mu }=\left(1,\vec{v}\right)$ is the four-vector velocity,
which we assume to be nearly light-speed. Only the lower order terms
of the correction were kept.

Thus, it is as if a cubic term is added to the dispersion relation.
However, this model differs from the one of ref.~\cite{Amelino} in the
sense that eq.~(\ref{cubic_eq_disp}) is sensible to the typical energy
of the product of the reaction, that is, to the process in
question. Also, there is a dependence on the geometry of the
propagation of the particle, through the term $s^{\mu }v_{\mu }$. In
addition, since the free theory is not altered by our approach, this
effective dispersion equation may only be used to study particle
reactions and not classical free-particle propagation.

\section{Quantum field theory}

To study the quantum aspects of the noncommutative model discussed in
sections II and III we consider the $\lambda \Phi ^{4}$ theory. The
action is given by
\begin{equation}
S=\int _{*}d^{n}x\left[{1 \over 2}\partial _{\mu }\Phi *\partial ^{\mu
  }\Phi +{m^{2} \over 2}\Phi *\Phi +{\lambda \over 4!}\Phi *\Phi *\Phi
  *\Phi \right]
\end{equation}
and can be evaluated in Fourier space to yield the corresponding
Feynman rules. The vertices are already calculated in
eq.~(\ref{nc_vertex}) and the propagators are the same as the
commutative ones. We shall consider the euclidean formulation.

The free propagator obeys the usual energy-momentum conservation.
However, since interactions do not, it is expected that the quantum
corrections to the propagator due to the self-interaction term in the
action will not obey the usual momentum conservation. Therefore, it is
of interest to compute the first order correction to the two-point
function, which is proportional to
\begin{equation}
\Gamma _{1}\left(p_{i},p_{f}\right)=-{1 \over 2}{\lambda \over 4!}{1
  \over p_{i}^{2}+m^{2}}{1 \over
  p_{f}^{2}+m^{2}}I\left(p_{i},p_{f}\right)
\end{equation}
with
\begin{equation}
I\left(p_{i},p_{f}\right)=\int {d^{n}q \over \left(2\pi
  \right)^{n}}{\sum _{\pi }\delta
  \left[p_{i}+p_{f}+V^{4}\left(p_{i},p_{f},q,-q\right)\right] \over
  q^{2}+m^{2}}\,,
\label{I}
\end{equation}
where the sum is computed over all $\pi $ permutations of the
arguments of $V^{4}$.

We proceed by evaluating this integral in the simple model
eq.~(\ref{nil2}).  From the several contributions to the above
integral, there are two which differ from the commutative case. They
are formally identical, and the relevant integrals are given by
\begin{equation}
J\left(p_{i},p_{f}\right)=\int {d^{n}q \over \left(2\pi \right)^{n}}{1
  \over q^{2}+m^{2}}\delta
[p_{i}+p_{f}+2V(q,k)]
\label{int_prop_corr}
\end{equation}
where $k=p_{i},p_{f}$ for each case.

Using a Schwinger parametrization and the usual Fourier representation
for the delta-function, the integrals are gaussian and yield
\begin{equation}
J\left(p_{i},p_{f}\right)={1 \over \left(4\pi \right)^{n}\sqrt{\det
    N}}{1 \over \left(p_{i}+p_{f}\right)\cdot N^{-1}\cdot
  \left(p_{i}+p_{f}\right)+m^{2}}\, ,
\label{prop_correctio}
\end{equation}
where
\begin{equation}
N_{\omega \lambda }\left(k\right)=k_{\nu }\Lambda _{\quad \omega
}^{\mu \nu }k_{\beta }\Lambda _{\mu \, \lambda }^{\, \beta }\, .
\label{Matrz_N}
\end{equation}

Notice that this matrix is singular for model eq.~(\ref{solution}),
but not in general. The fact that $\Lambda $ is nilpotent is not
important, as $\Lambda _{\quad \alpha }^{\mu \nu }$ is nilpotent only
regarding the indices $\mu $ and $\alpha $, and $N$ involves only
$\alpha $ type indices. Also, matrix $N$ is singular if $k=p_{i}=0$ or
$k=p_{f}=0$ and, hence, there is an IR divergence, which is usual in
noncommutative quantum field theories.

The integral
\begin{equation}
J_{C}\left(p_{i},p_{f}\right)=\delta \left(p_{i}+p_{f}\right)\int
{d^{n}q \over \left(2\pi \right)^{n}}{1 \over q^{2}+m^{2}}\, ,
\end{equation}
which is ultraviolet divergent for $n=4$, also arises from
eq.~(\ref{I}).  Thus one concludes that its regularization is still
required, although this is not necessary in
eq.~(\ref{int_prop_corr}). Hence, the UV renormalizability properties
at the one-loop approximation are not altered.

From the correction to the two-point function in
eq.~(\ref{prop_correctio}) two interesting features arise. First, the
conservation of momentum is lost, since the delta function $\delta
\left(p_{i}+p_{f}\right)$ is no longer present. Second, the correction
to the dispersion relation, which is given by the pole of
eq.~(\ref{prop_correctio}), manifests itself in a quite specific way,
involving the Lorentz algebra of the momentum vectors and matrix
$N^{-1}$. In fact, the poles of eq.~(\ref{prop_correctio}) suggest the
particle is subjected to a momentum-dependent metric given by
$N^{-1}$.

\section{Discussion and conclusions}

In this work we have presented a noncommutative field theory where the
coordinates have a Lie-algebra commutator as eq.~(\ref{intro}) with
nilpotent structure constants.  This breaks Lorentz as well as
translational invariance. Free theory is unchanged so the propagators
and the dispersion relations are not altered. The vertices show a new
energy-momentum law, which steems from the breaking of translational
invariance. The kinematical studies of such law where established in
particle decay physics and shown how it can be applied as a possible
explanation for the violation of the GZK cutoff, setting the
noncommutative mass scale at $M\approx 4\times 10^{19}eV$.  A link
between these kinematics and Lorentz-violating theories was
established, using a simplified model. However, there are well
definite differences between our approach and the ones usual discussed
in the literature (see ref.~\cite{Bertolami}), the most important one
being that an effective dispersion law always depends on the energy
and geometry of the processes in question.

It is tempting to speculate that our results have a bearing on the
other known astrophysical puzzles, namely the observation of high
energy photons, $\epsilon \approx 20\, TeV$, from far away sources and
the pion stability in extensive air showers (see ref.~\cite{Bertolami}
and references therein). Indeed, since both phenomena can be
understood via a cubic deformation in the relativistic dispersion
relation, so at pair creation through the process $\gamma +\gamma
_{IR\, bacground}\rightarrow e^{+}+e^{-}$ cannot occur and pion decay
into photons has a smaller width, it is plausible to assume that these
paradoxes can be explained in our model as well.

In the context of quantum field theory, it was shown that it is
possible to carry out explicit calculations regarding the first-order
correction to the two-point function in $\lambda \Phi ^{4}$
theory. The interaction terms violate momentum conservation and this
is expressed in the two-point function, where the usual delta function
structure $\delta \left(p_{i}+p_{f}\right)$ is lost. Even though the
noncommutative contributions are UV finite, usual commutative
integrals are still present and are UV divergent.  Thus, the UV
renormalization properties of one-loop calculations are not
altered. Also, in a strict sense, it is shown that free theory is
unchanged and so the propagators and the dispersion relations are not
altered. New IR divergences arise in the noncommutative corrections, a
feature which is shared with constant commutator noncommutative field
theories, known as UV/IR mixing. The poles of the noncommutative terms
indicate that there is a correction to the dispertion relation,
through the Lorentz algebra of matrix $N$, eq.~(\ref{Matrz_N}), which
seems to indicate that the particle satifies a dispersion relation
arising from a momentum-dependent metric, eq.~(\ref{Matrz_N}).

\acknowledgments

The authors would like thank Justin Conroy, Daniel Robbins, Naoki
Sasakura, Ricardo Schiappa and Edward Witten for their comments and
constructive remarks.

\end{document}